\documentclass[a4paper,11pt]{article}
\usepackage{pos}
\usepackage{physics}
\usepackage{slashed}
\usepackage{subcaption}

\title{
\vspace*{-0.75cm}
\begin{minipage}{\textwidth}
\begin{flushright}
\texttt{\footnotesize
PoS(LATTICE2024)304\\
ADP-25-3/T1265\\
DESY-25-020\\
Liverpool LTH 1395\\
}
\end{flushright}
\end{minipage}\\[15pt]
\vspace*{+0.75cm}
Transverse force distributions in the proton from lattice QCD}
\ShortTitle{Transverse force distributions in the proton from lattice QCD}

\author[a]{K.~U.~Can}
\author*[a]{J.~A.~Crawford}
\author[b]{R.~Horsley}
\author[c]{P.~E.~L.~Rakow}
\author[d]{G.~Schierholz}
\author[e]{H.~St\"uben}
\author[a]{R.~D.~Young}
\author[a]{J.~M.~Zanotti}

 \onbehalf{for the QCDSF collaboration}

\affiliation[a]{
CSSM, Department of Physics, The University of Adelaide, Adelaide SA, 5005, Australia
}
\affiliation[b]{School of Physics and Astronomy, University of Edinburgh, Edinburgh, EH9 3FD, UK}
\affiliation[c]{Theoretical Physics Division, Department of Mathematical Sciences, University of Liverpool, Liverpool L69 3BX, UK}
\affiliation[d]{Deutsches Elektronen-Synchrotron DESY, Notkestr. 85, 22607 Hamburg, Germany}
\affiliation[e]{Universit\"at Hamburg, Regionales Rechenzentrum, 20146 Hamburg, Germany}

\emailAdd{joshua.crawford@adelaide.edu.au}

\abstract{Single-spin asymmetries observed in polarised deep-inelastic scattering are important probes of hadron structure. The Sivers asymmetry provides information about the transverse momentum of the struck quark and can be related to final-state interactions. Understanding these asymmetries at the quark level has been the subject of much interest in QCD phenomenology. In this contribution, we present a lattice QCD calculation of the transverse spatial distribution of a colour-Lorentz force acting on the struck quark in a proton. Our lattice calculations employ $N_f = 2 + 1$ flavours of dynamical fermions at the SU(3) symmetric point across three lattice spacings. We determine a central, spin-independent confining force, as well as spin-dependent force distributions with local forces larger than the QCD string tension. These distributions offer a new, complimentary picture that underlies the Sivers asymmetry in transversely polarised deep-inelastic scattering.}

\FullConference{%
  The 41st International Symposium on Lattice Field Theory (LATTICE2024)\\
  28 July - 3 August 2024\\
  Liverpool, UK
}

\begin{document}
\maketitle

\section{Introduction}
Deep-inelastic scattering (DIS) provides a probe of the partonic structure of hadrons. Contributions to the DIS cross-section are categorised by their twist, which indicates the order in $Q^{-2}$ at which they enter the cross-section. Matrix elements of twist-two operators have physical interpretations within the parton model, however in special cases, this partonic intuition can be used to interpret matrix elements of higher-twist operators. One such case is matrix elements of certain twist-three operators having a semi-classical interpretation as a colour-Lorentz force acting on the struck quark \cite{Burkardt2013}. These forces become of interest in the context of asymmetries observed in polarised DIS experiments.
\par

In the following, we present a summary of our lattice QCD calculation of the distribution of this colour-Lorentz force in transverse impact parameter space. The full details of the calculation can be found in Ref. \cite{Crawford2024}. This involves extending computations of the forward matrix element $d_2$ to off-forward kinematics and computing three new form factors which encode these distributions. We show that these force distributions offer a complementary perspective on asymmetries observed in DIS experiments, and reveal large, non-trivial force structures that act within the proton.

\section{Single-Spin Asymmetries and Colour-Lorentz Forces}
DIS experiments which introduce a polarisation direction to the target hadron observe a surprising asymmetry in the azimuthal distribution of final states \cite{HERMES2007, HERMES2008, COMPASS2006}. In the context of transversely polarised semi-inclusive DIS, one such single-spin asymmetry is the Sivers asymmetry \cite{Sivers1991}. This asymmetry reveals that the transverse momentum of unpolarised quarks in a transversely polarised nucleon is not uniform, and that this asymmetric distribution of momentum is correlated with the nucleon spin \cite{Collins1992, Collins2002}. This is captured in the Sivers transverse momentum distribution (TMD), which allows for tomographic scanning of the nucleon in transverse momentum space. Alternatively, we compute matrix elements of twist-three operators which obtain physical interpretations as forces through the chromodynamic lensing framework \cite{Burkardt2003}. These forces then offer insights into the nature of the Sivers asymmetry. The approach used here is complementary to the TMD approach as we perform tomographic scanning of the nucleon in transverse impact parameter space, which is not the Fourier conjugate of transverse quark momentum.
\par

The hadronic contribution to the DIS cross section is contained in the hadronic tensor, which in turn, is parameterised by structure functions. When considering the spin-dependent structure functions $g_1(x,Q^2)$ and $g_2(x,Q^2)$, the $g_2$ structure function offers unique insights into baryon phenomenology as at leading order in $Q^{-2}$, it receives contributions from both twist-two and twist-three operators \cite{Jaffe1989, JaffeJi1990, Blumlein1996}. A particular case of twist-three matrix elements gaining partonic interpretations is the second moment of the twist-3 part of $g_2$, commonly denoted $d_2$ in the literature. There are a small number of lattice QCD calculations of $d_2$ in the literature \cite{Gockeler2000,Gockeler2005,Burger2022}. $g_2$ can be written as the sum of the twist-2 Wandzura-Wilczek contribution \cite{WandzuraWilczek1977} and a twist-3 term \cite{Cortes1991},
\begin{equation}
    g_2(x,Q^2) = g_2^{WW}(x,Q^2) + \Tilde{g}_2(x,Q^2).
\end{equation}
The second moment of the twist-three part of $g_2(x,Q^2)$ can be written in terms of a local light-cone matrix element through the operator product expansion,
\begin{equation}
\label{eq: forward twist 3 ME}
    3\int_{-1}^1 dx\, x^2 \Tilde{g}_2(x) = d_2 = \frac{1}{2mP^+ P^+ S^x}\mel{P,S}{\overline{q}(0)\gamma^+ g G^{+y} q(0)}{P,S}.
\end{equation}
An alternative interpretation of this twist-three matrix element is that it represents an average \textit{colour-Lorentz} force acting on the struck quark. By considering the local light-cone operator, we can observe the quark current, $\overline{q}(0)\gamma^+q(0)$, as well as the light-cone component of the gluon field strength tensor, $G^{+y}$. By converting this component of the gluon field strength tensor back to Cartesian coordinates and expressing these components in terms of chromo-electric and chromo-magnetic fields, this reveals the presence of a force analagous to the Lorentz force in electrodynamics,
\begin{equation}
    G^{+y} = \frac{1}{\sqrt{2}}\left(G^{0y} + G^{0z} \right) = -\frac{1}{\sqrt{2}}\bigg[ \vec{E}_c + \vec{v}\times \vec{B}_c\bigg]^{y} = -\frac{1}{\sqrt{2}}F^y,
\end{equation}
where $\vec{E}_c$ and $\vec{B}_c$ are the chromo-electric and chromo-magnetic fields respectively. Therefore, calculations of $d_2$ can provide information on the magnitude of the average colour-Lorentz force acting on the struck quark at the instant the virtual photon is absorbed \cite{Burkardt2013}.
\par

This idea of colour-Lorentz forces has been extended in Ref. \cite{Burkardt2019} to allow for the imaging of their distribution in transverse impact parameter space. Here, the local, forward matrix element in Equation \eqref{eq: forward twist 3 ME} is generalised to off-forward kinematics and parameterised in terms of form factors, 
\begin{multline}
\label{eq: T3 ME FF param}
    \mel{p',s'}{\overline{q}(0)\gamma^+ ig G^{+i}(0)q(0)}{p,s} = \overline{u}(p',s')\bigg[ P^+\Delta^i \gamma^+ \Phi_1(t) + MP^+i\sigma^{+i}\Phi_2(t) \\ + \frac{1}{M}P^+\Delta^ii\sigma^{+\Delta}\Phi_3(t) \bigg]u(p,s),
\end{multline}
where $P^\mu = (p'+p)^\mu/2$, $\Delta^\mu = (p'-p)^\mu$, $t = -\Delta^2$ and $\sigma^{\mu\Delta} = \sigma^{\mu\nu}\Delta_\nu$. Furthermore, we consider the zero-skewness limit which implies that $\Delta^+ = 0$ such that density distributions in impact parameter space obtain a probabilistic interpretation \cite{Burkardt2000}. Computing the 2D Fourier transforms of the form factors then reveals the distribution of this colour-Lorentz force in transverse impact parameter space.

\section{Calculation Details}
Our calculations make use of three gauge ensembles generated by the QCDSF collaboration \cite{Bietenholz2011}, which employ fermions described by a stout-smeared non-perturbatively $\mathcal{O}(a)$-improved Wilson action \cite{Cundy2009} and a tree-level Symanzik improved gluon action. We refer the reader to Ref. \cite{Crawford2024} for the full lattice details. Two randomly-generated source locations are used per configuration. We compute the required hadronic matrix elements through ratios of 3- and 2-pt functions,
\begin{equation}
    \mathcal{R} = \frac{C_{3pt}(\mathbf{q},\tau;\mathbf{p}',t;\mathcal{O})}{C_{2pt}(\mathbf{p}',t)}\sqrt{\frac{C_{2pt}(\mathbf{p}',t)C_{2pt}(\mathbf{p}',\tau)C_{2pt}(\mathbf{p},t-\tau)}{C_{2pt}(\mathbf{p},t)C_{2pt}(\mathbf{p},\tau)C_{2pt}(\mathbf{p}',t-\tau)}},
\end{equation}
where $t$ is the source-sink temporal separation, $\tau$ is the operator insertion time slice, $\mathbf{q} = \mathbf{p}' - \mathbf{p}$ is the transfer momentum. In the large Euclidean time limit, this ratio is proportional to the matrix element of the inserted operator $\mathcal{O}$. However, to account for excited state contamination, we fit a two-state model for the 3- and 2-pt functions and extract the ground state contribution.
\par

Due to the limited symmetries of the hypercubic lattice, operators which transform under the same irreducible representation of the hypercubic group $H(4)$ can mix with lower dimensional operators \cite{Baake1981}. Our operator of interest,
\begin{equation}
    \mathcal{O}^{[5]}_{[i\{j]4\}}=-\frac{1}{4}\overline{q}(0)\gamma_{[i}\gamma_5 \overleftrightarrow{D}_{\{j]}\overleftrightarrow{D}_{4\}}q(0),
\end{equation}
has non-vanishing mixing with the operator, 
\begin{equation}
    \mathcal{O}^\sigma_{[i\{j]4\}} = \frac{i}{12}\overline{q}(0)\left(\sigma_{jk}\overleftrightarrow{D}_j - \sigma_{4k}\overleftrightarrow{D}_4 \right)q(0),
\end{equation}
where $i,j,k \in \{1,2,3 \}$, $i\neq j\neq k$, $\overleftrightarrow{D}=\frac{1}{2}(\overrightarrow{D}-\overleftarrow{D})$ is the symmetrised covariant derivative, and $\{...\}$ ($[...]$) denotes (anti-)symmetrisation of indices. The operator $\mathcal{O}^{[5]}_{[i\{j]4\}}$ can be related to the operator in Equation \eqref{eq: T3 ME FF param} by using the identities $\overleftrightarrow{D}_\mu = \frac{1}{2}\left\{ \gamma_\mu, \slashed{D}\right\}$, $[D_\mu,D_\nu] = g G_{\mu\nu}$ and the QCD equations of motion $\slashed{D}q = \overline{q}\slashed{D} = 0$ \cite{Shuryak1982}. We compute matrix elements of both operators, and determine the degree of mixing between the two through the RI$^\prime$-MOM scheme, using the methods outlined in Refs \cite{Martinelli1994, Gockeler1998}. The renormalised operator $\mathcal{O}^{[5]}_R$, omitting indices, at some scale $\mu$ is then
\begin{equation}
    \mathcal{O}^{[5]}_R(\mu) = Z^{[5]}(a\mu)\left(\mathcal{O}^{[5]}(a) + \frac{1}{a}\frac{Z^\sigma(a\mu)}{Z^{[5]}(a\mu)}\mathcal{O}^\sigma(a) \right).
\end{equation}
The overall renormalisation constant $Z^{[5]}(a\mu)$ is calculated using the method detailed in Ref. \cite{Burger2022}.

\section{Results and Discussion}
\begin{figure}
    \centering
    \begin{subfigure}[t]{0.49\textwidth}
        \includegraphics[width=\linewidth]{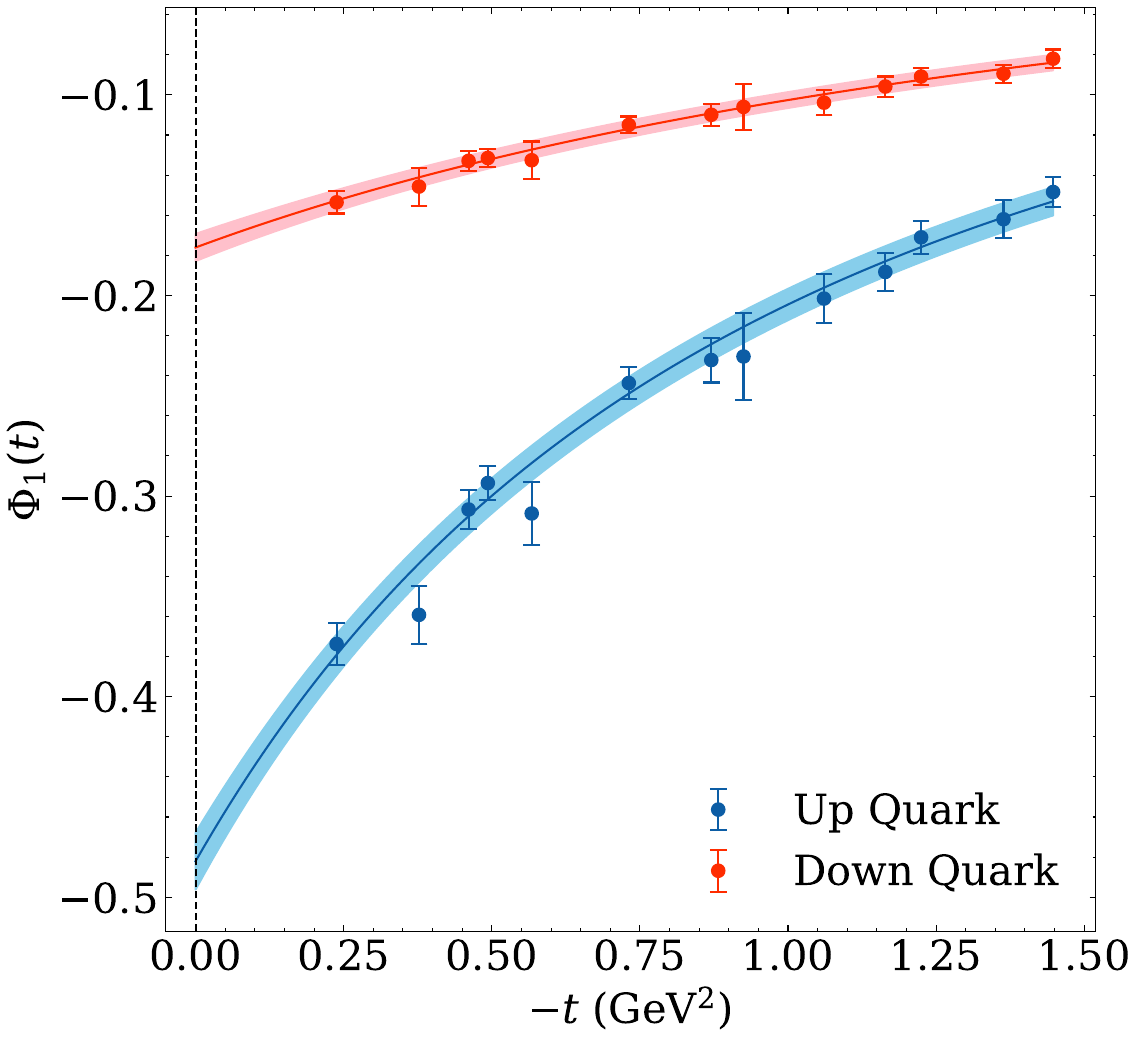}
        \caption{Renormalised $\Phi_1$ form factor results.}
        \label{fig: phi1 FF}
    \end{subfigure}
    \begin{subfigure}[t]{0.49\textwidth}
        \includegraphics[width=\linewidth]{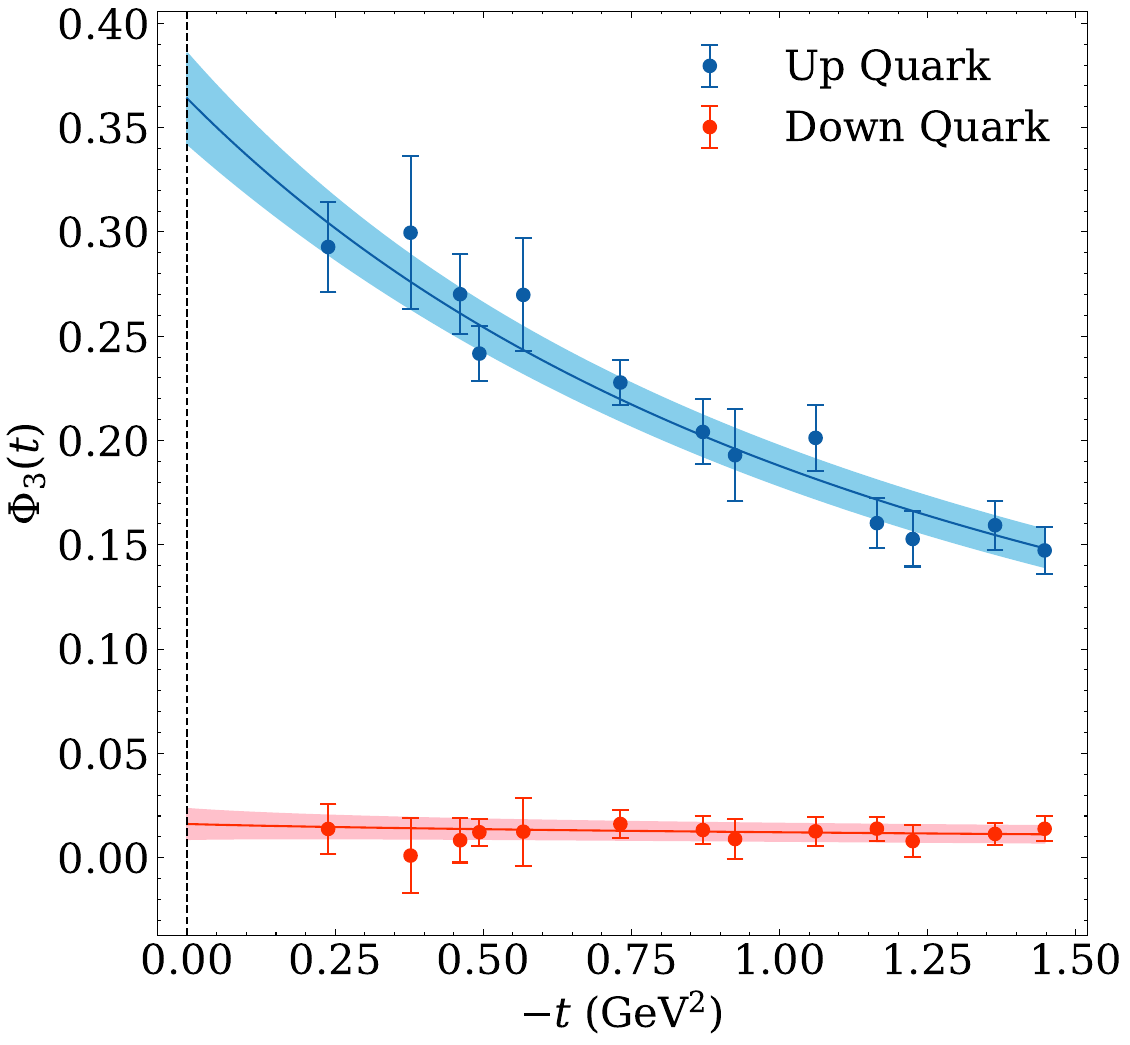}
        \caption{Renormalised $\Phi_3$ form factor results.}
        \label{fig: phi3 FF}
    \end{subfigure}
    \caption{Form factors computed at our finest lattice spacing with corresponding dipole fits.}
    \label{fig: combined FF fig}
\end{figure}

In Figure \ref{fig: phi1 FF}, we show the lattice results for the $\Phi_1$ form factor for both the up and down quarks at our finest lattice spacing, $a = 0.052$ fm. Results across our full set of lattices can be found in Ref. \cite{Crawford2024}. Strong signals with small statistical uncertainties for both quark flavours is shown. The $\Phi_1$ form factor is negative for both quark flavours, which when taking the 2D Fourier transform, is indicative of a universally attractive force. Furthermore, we note that the magnitude of the up quark $\Phi_1$ form factor is roughly double in size relative to the down quark, which is consistent with quark counting expectations. $\Phi_1$ is fit using a dipole ansatz, 
\begin{equation}
    \label{eq: dipole}
    \Phi_i(t) = \frac{\Phi_i(0)}{\left(1 -\frac{t}{\Lambda_i^2}  \right)^2},
\end{equation}
where $\Lambda_i$ is the dipole mass. Both quark flavours are well modelled by the dipole fit.
\par

Figure \ref{fig: phi3 FF} shows the lattice results for the $\Phi_3$ form factor for both quark flavours. Whilst the up quark shows a strong, non-zero signal, it proved difficult to isolate a non-zero signal for the down quark. A possible explanation for this is through a simple quark-diquark interpretation, where the down quark is bound within a scalar diquark and hence its contributions to any spin-dependent observables is suppressed. This hypothesis can be tested through models which emphasise the diquark scenario, such as those used in Refs. \cite{Cloet2012, Wang2024}.  The up quark results are well described by the dipole model from Equation \eqref{eq: dipole}, however the down quark fit needs to be reconstructed from fits to the isovector $(u-d)$ and isoscalar $(u+d)$ combinations.
\par

By taking a 2D Fourier transform of the form factors, we obtain a visualisation of the colour-Lorentz force distributions in the transverse plane,
\begin{equation}
    \mathcal{F}_{s's}^j = \int \frac{d^2\Delta_\perp}{(2\pi)^2}e^{-i\mathbf{b}\cdot\mathbf{\Delta}_\perp}F_{s's}^j(\mathbf{\Delta}_\perp),
\end{equation}
where,
\begin{equation}
\begin{split}
\label{eq: FF decomp}
    F_{s's}^j(\mathbf{\Delta}_\perp) &= \frac{i}{\sqrt{2}P^+}\mel{p^+,\frac{\Delta_\perp}{2},s'}{\overline{q}(0)\gamma^+ ig G^{+j}(0)q(0)}{p^+,-\frac{\Delta_\perp}{2},s},\\
    &= \overline{u}(p',s')\bigg[ P^+\Delta^i \gamma^+ \Phi_1(t) + MP^+i\sigma^{+i}\Phi_2(t)  + \frac{1}{M}P^+\Delta^ii\sigma^{+\Delta}\Phi_3(t) \bigg]u(p,s).
\end{split}
\end{equation}
Here, $j =x,y$ represents the transverse index, $s'$, $s$ denote the nucleon polarisation and $\mathbf{\Delta}_\perp$ is the transverse momentum transferred to the nucleon. In the following, we denote the contribution to the total force from the $\Phi_i$ form factor as $\mathcal{F}_i$.
\par

\begin{figure}
    \centering
    \includegraphics[width=0.8\linewidth]{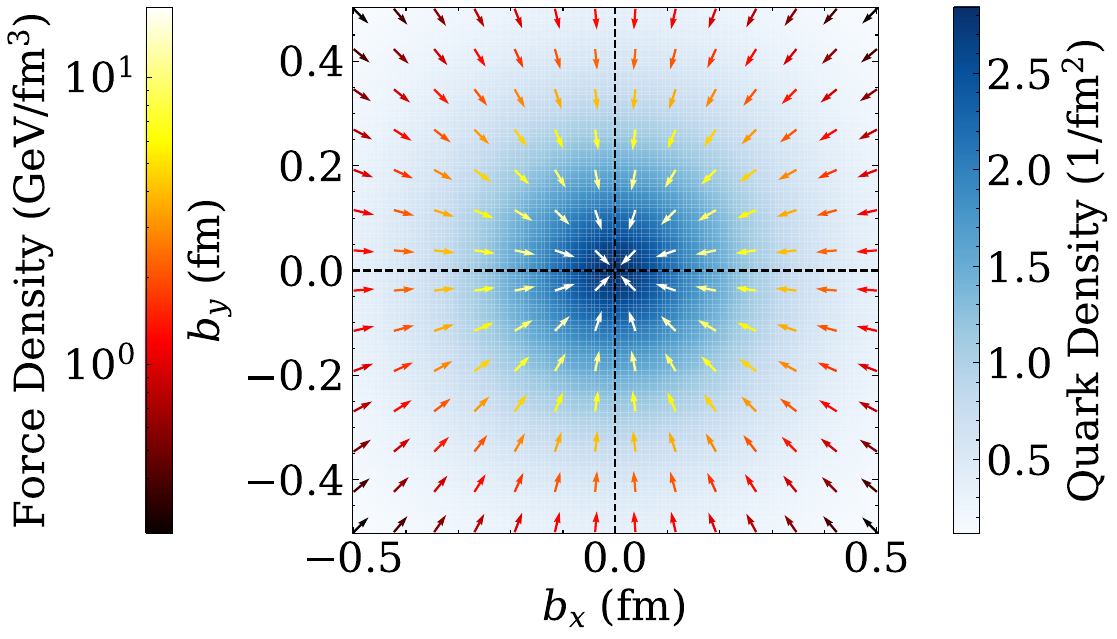}
    \caption{Distribution of the colour-Lorentz force acting on an unpolarised up quark in an unpolarised proton, indicated by the vector field. Vectors have unit length and their magnitude is indicated by the colour bar. The vector field is superimposed on the density distribution for the up quark in an unpolarised proton.}
    \label{fig: unpol force distribution}
\end{figure}

\sloppy The first term in the form factor decomposition, corresponding to the Dirac bilinear $\overline{u}(p',s')\gamma^+ u(p,s)$, is diagonal in the nucleon spins and therefore is not sensitive to the nucleon polarisation. As such, the Fourier transform of this term describes the colour-Lorentz force acting on unpolarised quarks in an unpolarised proton. This Fourier transform is shown in Figure \ref{fig: unpol force distribution}, where we have taken the Fourier transform of the dipole fit to $\Phi_1$. The force is shown to be attractive for all regions of impact parameter space, consistent with our understanding of confinement. The vectors shown in the figure are weighted by the corresponding quark density, and so to estimate the magnitude of these colour-Lorentz forces, we divide out the quark density dependence. The quark density is computed through 2D Fourier transforms of the electromagnetic form factors $F_1(t)$ and $F_2(t)$ using the procedure outlined in Refs. \cite{QCDSF2006, Diehl2005}. The quark density distributions are plotted alongside the force distributions to aid in visualisation.
\par

\begin{figure}
    \centering
    \includegraphics[width=0.8\linewidth]{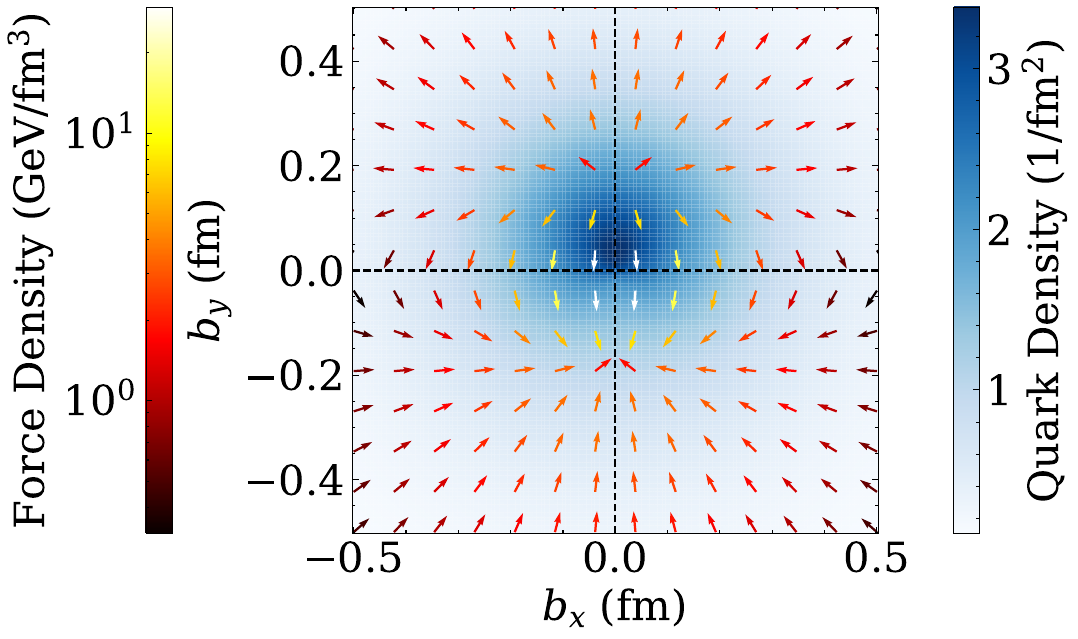}
    \caption{Distribution of the colour-Lorentz force acting on an unpolarised up quark in a proton polarised in the $\hat{x}$-direction, superimposed on the density distribution for the up quark in the polarised proton. Image conventions are identical to Figure \ref{fig: unpol force distribution}.}
    \label{fig: pol force distribution}
\end{figure}

The second two terms in Equation \eqref{eq: FF decomp} are proportional to the Dirac bilinear $\overline{u}(p',s')i\sigma^{+\mu}u(p,s)$, and require a nucleon spin-flip, hence they are sensitive to the nucleon transverse polarisation. Therefore a 2D Fourier transform of these terms describes the colour-Lorentz forces acting on unpolarised quarks in a polarised nucleon. Figure \ref{fig: pol force distribution} shows the summed contributions of the Fourier transforms of the $\Phi_2$ and $\Phi_3$ terms, when using the dipole fit for the form factors and considering a proton polarised in the $\hat{x}$-direction. The up quark density in an $\hat{x}$-polarised proton is similarly shown alongside to assist in visualisation. The resulting force distribution has a resemblance to a magnetic dipole from classical electromagnetism. Of note is the large magnitude of forces acting in the $-\hat{y}$ direction in the region of largest up quark density. These suggest that the struck up quark experiences a large force pulling it back towards the shattered remnants of the proton core as it leaves the bound state. This picture is consistent with the Sivers asymmetry, which shows an asymmetric distribution of final states opposite to that of the quark density asymmetry. Curiously, at large peripheral distances in the $+\hat{y}$ direction, the force is repulsive. However, given the low quark density in this region, it remains a challenge on how to reveal this effect (if at all possible) in an experimental cross-section or asymmetry.
\par

\begin{figure}
    \centering
    \begin{subfigure}[t]{0.49\textwidth}
        \includegraphics[width=\linewidth]{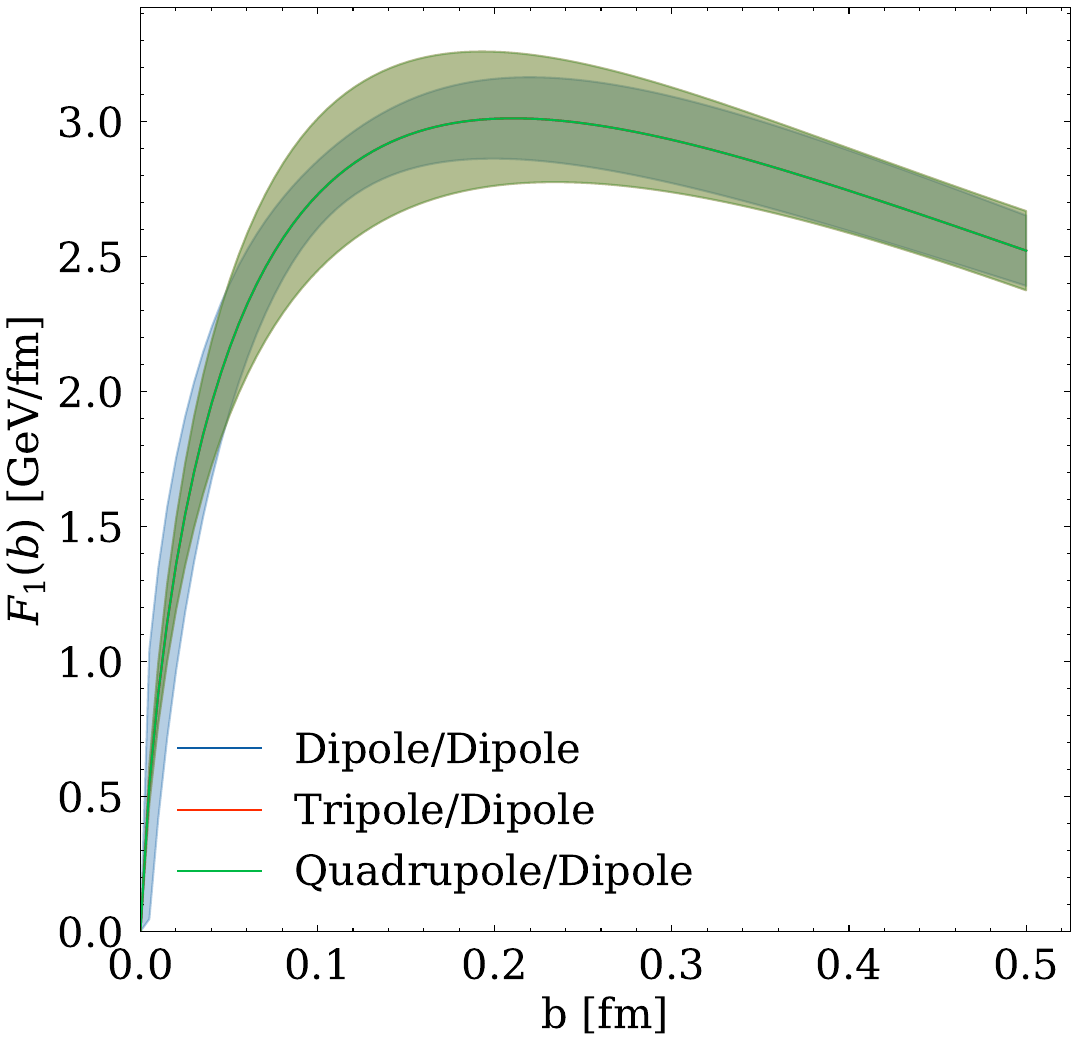}
        \caption{Colour-Lorentz force magnitude in the radial direction for an unpolarised up quark in an unpolarised proton.}
        \label{fig: unpolarised force magnitude}
    \end{subfigure}
    \begin{subfigure}[t]{0.49\textwidth}
        \includegraphics[width=\linewidth]{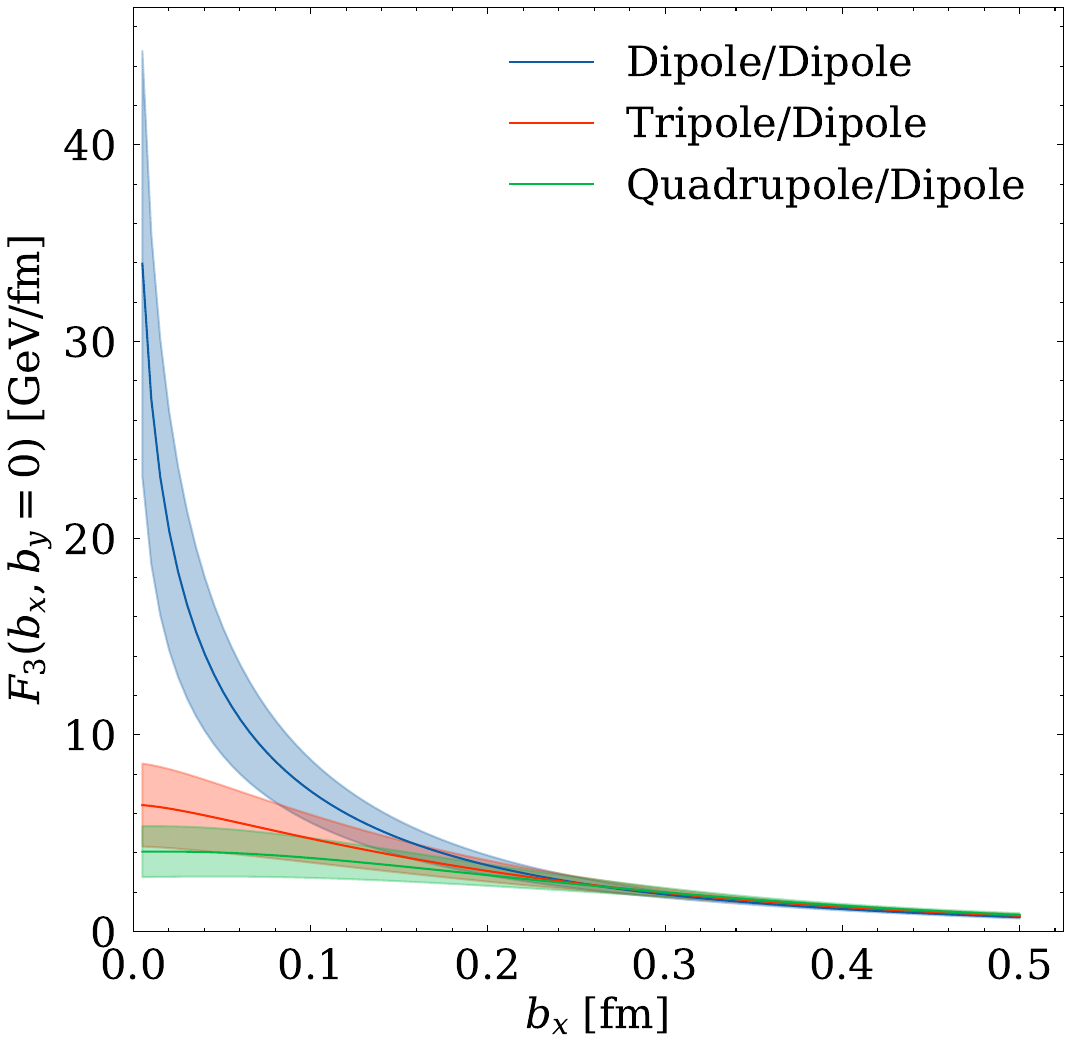}
        \caption{Colour-Lorentz force magnitude along the $\hat{x}$ axis for an unpolarised up quark in a polarised proton.}
        \label{fig: polarised force magnitude}
    \end{subfigure}
    \caption{Density-removed force profiles.}
    \label{fig: density removed force profiles}
\end{figure}

By assuming that the density-weighted force factorises in impact parameter space as
\begin{equation}
    \mathcal{F}^i_{s',s}(\mathbf{b}_\perp) = \rho_{s',s}(\mathbf{b}_\perp)F^i_{s',s}(\mathbf{b}_\perp),
\end{equation}
where $\rho_{s',s}(\mathbf{b}_\perp)$ is the quark density and $F^i_{s',s}(\mathbf{b}_\perp)$ is the unweighted colour-Lorentz force, we make an estimate of the force magnitudes by dividing out the quark density dependence for both the polarised and unpolarised cases. We also vary the functional form used to model the form factors as several different orders of multipoles,
\begin{equation}
    \Phi_i(t) = \frac{\Phi_i(0)}{\left(1 - \frac{t}{\Lambda_i^2} \right)^n},
\end{equation}
where $n = 2,3,4$, to assess the model dependence of the results. We make use of a dipole fit for the quark density distribution in all cases. The resulting force profiles are shown in Figure \ref{fig: density removed force profiles}. In Figure \ref{fig: unpolarised force magnitude}, we show the force magnitude acting on an unpolarised up quark in an unpolarised proton. We find that the colour-Lorentz force vanishes at the origin, before rising to a maximum of 3 GeV/fm at approximately 0.15 fm, and remains approximately constant for larger separations. We see consistent results between the three different multipole models used. These results are consistent with the expectations from the static quark potential, which predicts a constant restoring force at large distances. However, this local colour-Lorentz force is nearly three times as large as the phenomenological value for the QCD string tension, which defines the average force scale in QCD.
\par

In Figure \ref{fig: polarised force magnitude}, we show the force magnitude acting on an unpolarised up quark in a proton polarised in the $\hat{x}$-direction. Due to the non-trivial structure of the force field, we examine the force magnitude along the $\hat{x}$ axis. We note the diverging behaviour of the force models close to the origin, which indicates that further results for the $\Phi_3$ form factor at larger momentum are required to better constrain the model dependence. The force magnitude is essentially model independent at $\sim$ 0.25 fm, where the force magnitude has decreased to $\sim$ 3 GeV/fm. 

\section{Summary and Conclusion}
In this work, we have computed three new form factors of off-forward twist-three matrix elements using lattice QCD and determined the resulting distribution of colour-Lorentz forces in 2D impact parameter space. Strong signals were found for the up quark across the form factors, however the down quark was found to be suppressed in the polarised form factors. Expanding the momentum range on which the form factors are calculated will allow for a more rigorous model dependence study and less uncertainty in the force magnitude close to the origin. Computing the 2D Fourier transform of these form factors revealed distributions of large, local forces acting on the struck quark. To connect with phenomenology, these forces were interpreted in the context of the Sivers asymmetry, which provided a complementary perspective on the origin of the asymmetry. This study lays the ground work for further studies of the transverse distribution of these forces and the development of simple and intuitive visual representations of the origin of these asymmetries.

\acknowledgments
The authors would like to thank Matthias Burkardt for many useful discussions. The numerical configuration generation (using the BQCD lattice QCD program \cite{Haar2017}) and data analysis using the \texttt{CHROMA} software package \cite{Chroma2005}. Calculations were performed using the Cambridge Service for Data Driven Discovery (CSD3), the Gauss Centre for Supercomputing (GCS) supercomputers JUQUEEN and JUWELS (John von Neumann Institute for Computing, NIC, J\"ulich, Germany), resources provided by the North-German Supercomputer Alliance (HLRN), the National Computer Infrastructure (NCI National Facility in Canberra, Australia supported by the Australian Commonwealth Government), the Pawsey Supercomputing Centre, which is supported by the Australian Government and the Government of Western Australia and the Phoenix HPC service (University of Adelaide). JAC is supported by an Australian Government Research Training Program (RTP) Scholarship. RH is supported by STFC through the grant ST/X000494/1. KUC, RDY and JMZ are supported by the Australian Research Council grant DP220103098 and DP240102839. For the purpose of open access, the authors have applied a Creative Commons Attribution (CC BY) licence to any author accepted manuscript version arising from this submission.

\bibliographystyle{JHEP}
\bibliography{refs}

\end{document}